\documentclass[aps,prl,twocolumn,showpacs]{revtex4-1}
\usepackage[ansinew]{inputenc}
\usepackage{graphicx}
\usepackage{textcomp}
\usepackage{hyperref}

\newcommand{\ket}[1]{|#1\rangle} 

\begin{document}

\title{Nondestructive Detection of an Optical Photon}
\author{Andreas~Reiserer}
\author{Stephan~Ritter}
\email{stephan.ritter@mpq.mpg.de}
\author{Gerhard~Rempe}

\affiliation{Max-Planck-Institut f\"ur Quantenoptik, Hans-Kopfermann-Strasse 1, 85748 Garching, Germany}

\begin{abstract}
All optical detectors to date annihilate photons upon detection, thus excluding repeated measurements. Here, we demonstrate a robust photon detection scheme which does not rely on absorption. Instead, an incoming photon is reflected off an optical resonator containing a single atom prepared in a superposition of two states. The reflection toggles the superposition phase which is then measured to trace the photon. Characterizing the device with faint laser pulses, a single-photon detection efficiency of 74\% and a survival probability of 66\% is achieved. The efficiency can be further increased by observing the photon repeatedly. The large single-photon nonlinearity of the experiment should enable the development of photonic quantum gates and the preparation of novel quantum states of light.
\end{abstract}

\maketitle

More than a century ago, Planck's idea of a quantized energy exchange between light and matter and Einstein's conclusion that a light beam consists of a stream of particles have revolutionized our view of the world. The explanation of the photoelectric effect in terms of a photon-absorption process is the basis of the theoretical description of light with normally-ordered photon creation and annihilation operators \cite{glauber_quantum_1963,mandel_optical_1995}. The picture of photon detection as a destructive process has been confirmed experimentally ever since. Nondestructive detection \cite{braginsky_quantum_1996}, namely the ability to watch individual photons fly by, has until now been an unaccomplished ``ultimate goal'' \cite{grangier_quantum_1998} of optical measurements.

Nondestructive detection has two major implications. First, a single photon can be detected more than once. Thus, concatenating several devices improves the detection efficiency of single photons. Second, nondestructive detection can serve as a herald that signals the presence of a photon without affecting its other degrees of freedom, like its temporal shape or its polarization. This is in stark contrast to absorbing detectors, where the quantum state of the photon is projected and therefore lost. Both implications are of great importance for rapidly evolving research fields such as quantum measurement \cite{wiseman_quantum_2010}, optical quantum computing \cite{obrien_optical_2007}, and quantum communication and networking \cite{gisin_quantum_2007,kimble_quantum_2008}.

The interaction mechanism \cite{duan_scalable_2004} we implement is based on the principles of cavity quantum electrodynamics, remarkably robust, and applicable to many different physical systems. It allows one to nondestructively detect propagating optical photons and thus to complement experiments with microwave fields trapped in superconducting resonators \cite{nogues_seeing_1999,guerlin_progressive_2007,johnson_quantum_2010}. To this end, a faint laser pulse is reflected off a resonant cavity in which a trapped atom has been prepared in a superposition of two internal states. The cavity induces strong coupling between the light pulse and the atom in one of the atomic states, but not the other. This leads to a phase flip of the atomic superposition state upon reflection of a photon. Subsequent readout of the atomic phase thus makes it possible to detect a photon without absorbing it.

\begin{figure}
\includegraphics[width=1.0\columnwidth]{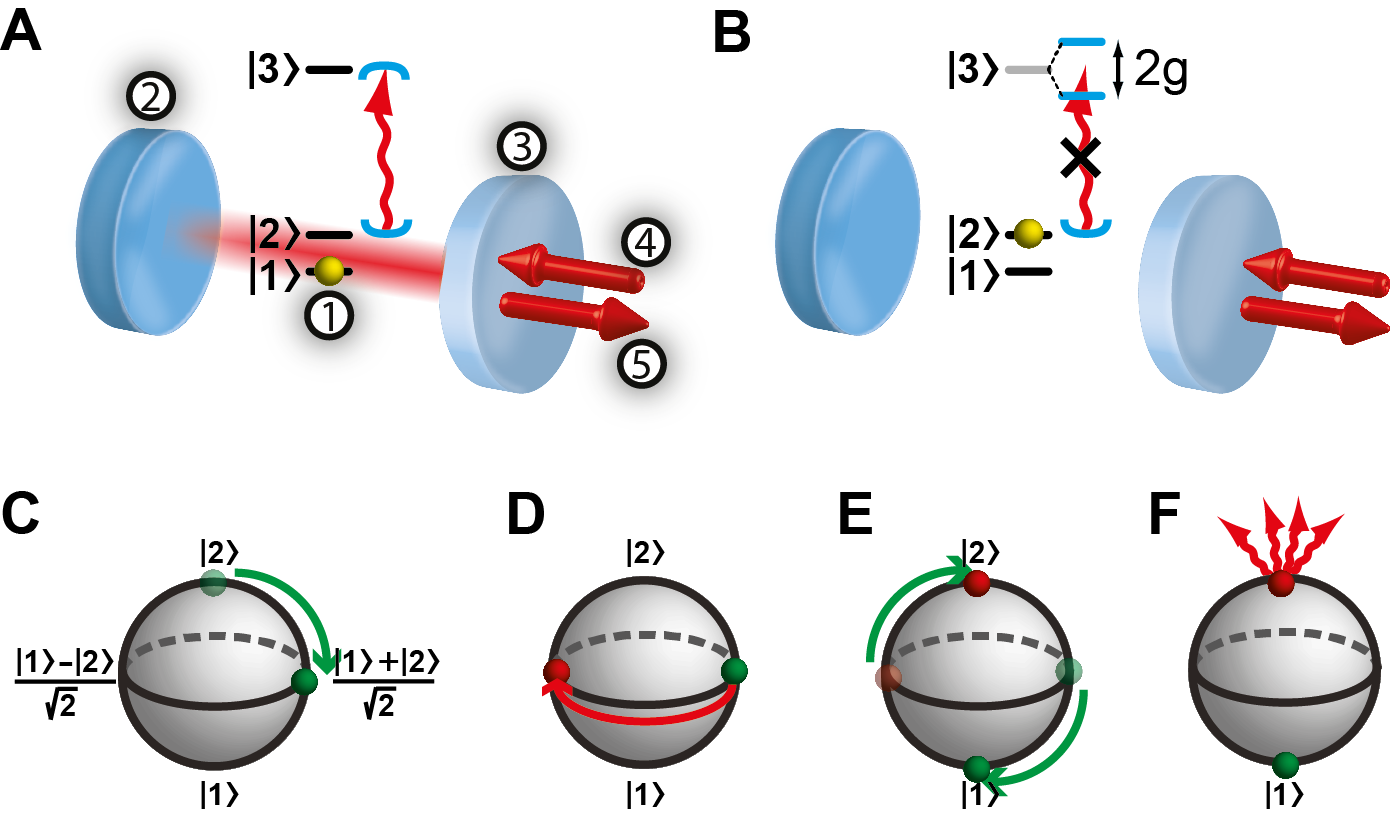}
\caption{\label{fig:setup}
\textbf{Nondestructive photon detection.} (\textbf{A,B}) Sketch of the setup and atomic level scheme. A single atom, (1), is trapped in an optical cavity that consists of a high-reflector, (2), and a coupling mirror, (3). A resonant photon is impinging on, (4), and reflected off, (5), the cavity. (\textbf{A}) If the atom is in state $\ket{1}_a$, the photon (red curly arrow) enters the cavity (blue semicircles) before being reflected. In this process, the combined atom-photon state acquires a phase shift of $\pi$. (\textbf{B}) If the atom is in $\ket{2}_a$, the strong coupling on the $\ket{2}_a \leftrightarrow \ket{3}_a$ transition leads to a normal-mode splitting of $2g$, such that the photon cannot enter the cavity and is directly reflected without a phase shift.
(\textbf{C to F}) Procedure to measure whether a photon has been reflected. (\textbf{C}) The atomic state, visualized on the Bloch sphere, is prepared in the superposition state $\frac{1}{\sqrt{2}}(\ket{1}_a+\ket{2}_a)$. (\textbf{D}) If a photon impinges, the atomic state is flipped to $\frac{1}{\sqrt{2}}(\ket{1}_a-\ket{2}_a)$. (\textbf{E}) The atomic state is rotated by $\frac{\pi}{2}$. (\textbf{F}) Fluorescence detection is used to discriminate between the states $\ket{1}_a$ and $\ket{2}_a$.
}
\end{figure}

A detailed theoretical treatment of the atom-photon interaction mechanism is given in \cite{duan_scalable_2004,cho_generation_2005}. For an intuitive explanation, consider a three-level atom in a single-sided cavity (Fig.\,\ref{fig:setup}A), where one of the mirrors is perfectly reflecting and the small transmission of the other mirror allows for in- and outcoupling of light. The cavity is thus overcoupled, and resonant with the transition between the atomic states $\ket{2}_a$ and $\ket{3}_a$. A photon, resonant with the empty cavity, is impinging onto the transmitting mirror. If the atom is in the state $\ket{1}_a$, it will not interact with the photon because any transition is far detuned. Thus, the photon will enter the cavity before being reflected. If, however, the atom is in $\ket{2}_a$ (Fig.\,\ref{fig:setup}B), the strong atom-photon coupling leads to a normal-mode splitting, such that the photon is reflected without entering the cavity. In this case, atom and photon were never in the same place. Nevertheless, the photon has left a trace in the state of the atom: When light is reflected from a resonant cavity, it experiences a phase shift of $\pi$, while there is no phase shift in the strongly coupled case. When the impinging photon is denoted by the state $\ket{1}_p$, we thus find $\ket{2}_a\ket{1}_p \rightarrow \ket{2}_a\ket{1}_p$, whereas $\ket{1}_a\ket{1}_p \rightarrow \mathrm{e}^{i\pi}\ket{1}_a\ket{1}_p=-\ket{1}_a\ket{1}_p$.

To use this conditional phase shift for nondestructive photon detection, the atom is prepared in the superposition state $\frac{1}{\sqrt{2}}(\ket{1}_a+\ket{2}_a)$ (Fig.\,\ref{fig:setup}C). If there is no impinging photon, the atomic state remains unchanged (Fig.\,\ref{fig:setup}D, green filled circle). If, however, a photon is reflected, the atomic state becomes (omitting a global phase) $\frac{1}{\sqrt{2}}(\ket{1}_a+\ket{2}_a)\ket{1}_p \rightarrow \frac{1}{\sqrt{2}}(\ket{1}_a-\ket{2}_a)\ket{1}_p$ (red arrow and red filled circle). To measure this phase flip, a $\pi/2$ rotation maps the atomic state $\frac{1}{\sqrt{2}}(\ket{1}_a+\ket{2}_a)$ onto $\ket{1}_a$, while $\frac{1}{\sqrt{2}}(\ket{1}_a-\ket{2}_a)$ is rotated to $\ket{2}_a$ (Fig.\,\ref{fig:setup}E). Subsequently, cavity-enhanced fluorescence state detection \cite{bochmann_lossless_2010} is used to discriminate between the atomic states $\ket{1}_a$ and $\ket{2}_a$ (Fig.\,\ref{fig:setup}F, \cite{som}). Note that two photons in the input pulse lead to a phase shift of $\mathrm{e}^{i 2\pi}=1$. The used sequence therefore measures the odd-even parity of the photon number. As long as the average photon number per measurement interval is much smaller than one, only zero or one photon events are present and the detection result is unambiguous.

In our setup \cite{reiserer_ground-state_2013}, a single $^{87}$Rb atom is trapped in a three-dimensional optical lattice at the center of a Fabry-Perot resonator. The coupling mirror has a transmission of 95\,ppm, which is large compared to the transmission of the high-reflector and the scattering and absorption losses (8\,ppm). The cavity field decay rate is $\kappa=2\pi\times2.5$\,MHz, the atomic dipole decay rate is $\gamma=2\pi\times3$\,MHz, and the measured atom-cavity coupling constant on the $\ket{2}_a \leftrightarrow \ket{3}_a$ transition is $g=2\pi\times6.7$\,MHz \cite{reiserer_ground-state_2013}. Thus, the system operates in the strong-coupling regime of cavity quantum electrodynamics.

\begin{figure}
\includegraphics[width=1.0\columnwidth]{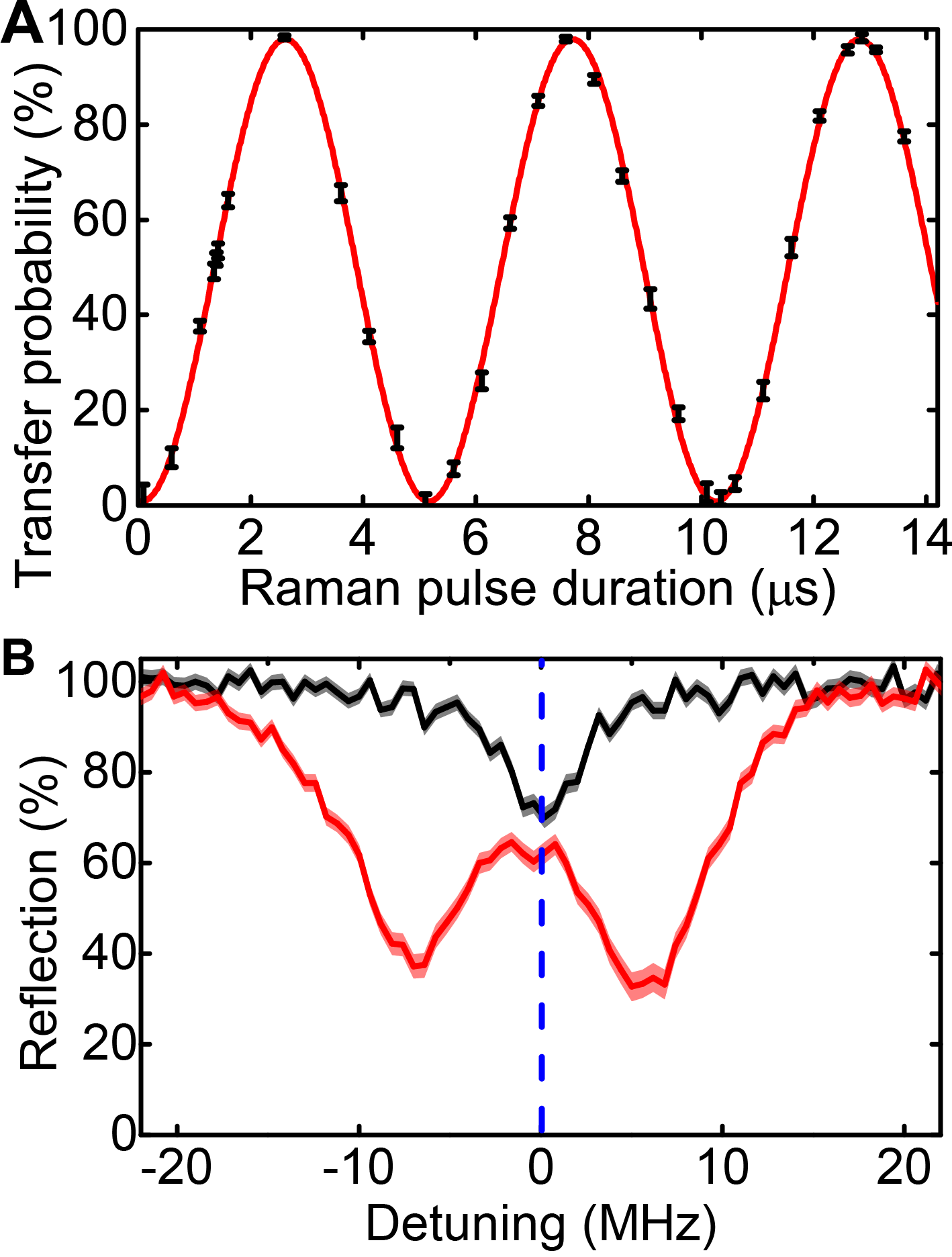}
\caption{\label{fig:IndividualSteps}
\textbf{Atomic state manipulation and cavity reflection spectrum.} (\textbf{A}) Rabi oscillations of the atomic population when the atom is prepared in $\ket{2}_a$ and two Raman laser beams are applied for a variable duration. The red fit curve gives a visibility of 97\%.
(\textbf{B}) Reflection off the atom-cavity system as a function of probe laser frequency with the atom in the strongly coupled state $\ket{2}_a$ (red) or in the uncoupled state $\ket{1}_a$ (black). The statistical standard error is given by the thickness of the lines.
}
\end{figure}

We first demonstrate that we can accurately prepare, control and read out the atomic state. The atom is initialized in the state $\ket{2}_a$ by optical pumping and the levels $\ket{2}_a$ and $\ket{1}_a$ are coupled using a pair of Raman lasers \cite{som}. To characterize this coupling, the Raman beams were applied for a variable duration and the population in $\ket{2}_a$ was measured \cite{bochmann_lossless_2010,som}. Observing Rabi oscillations (Fig.\,\ref{fig:IndividualSteps}A) with a visibility of $97\%$ represents an upper bound for the quality of our state preparation, rotation and readout process.

Strong coupling between the atom and impinging light is demonstrated by measuring the reflection of the system with the atom prepared in $\ket{2}_a$ as a function of the probe light frequency (red data in Fig.\,\ref{fig:IndividualSteps}B). The observed normal-mode splitting testifies to the strong coupling. On resonance, 62(2)\% of the impinging photons are reflected. With increasing coupling strength, this value is expected to approach unity. When the atom is prepared in the uncoupled state $\ket{1}_a$, 70(2)\% of the incoming light is reflected on resonance (black data in Fig.\,\ref{fig:IndividualSteps}B). The missing 30\% are either transmitted through the high-reflector or lost via scattering or absorption, in good agreement with input-output theory calculations \cite{walls_quantum_2008} using the independently measured mirror parameters.

\begin{figure}
\includegraphics[width=1.0\columnwidth]{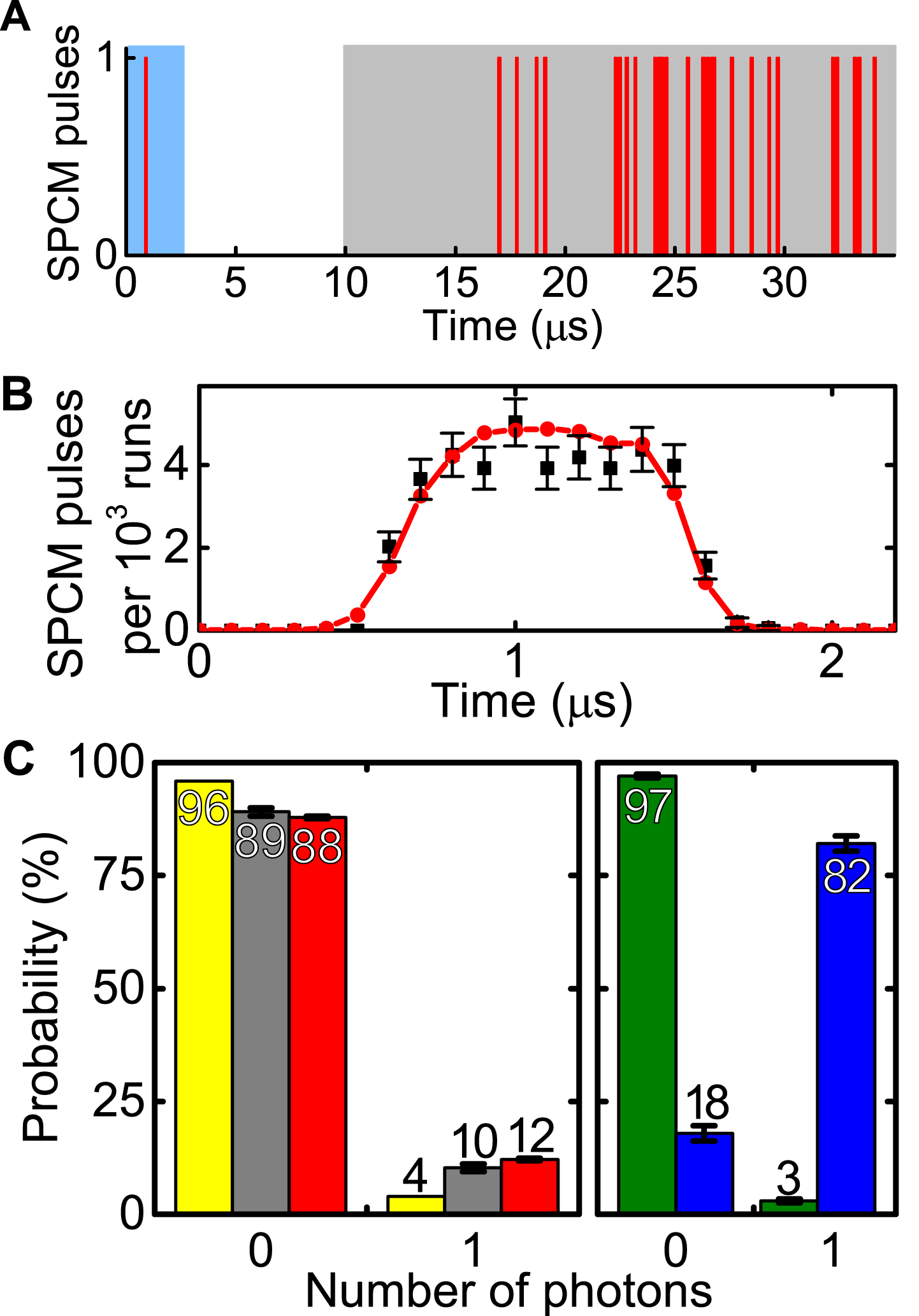}
\caption{\label{fig:QND}
\textbf{Experimental results.} (\textbf{A}) Typical trace of an experimental run. A photon (red bar) impinging in the trigger interval (blue area) leads to the emission of many photons in the readout interval (grey area). When the input pulse is blocked, no photons are detected in both intervals.
(\textbf{B}) Temporal envelope of the reflected photon pulse when an atom is present (black squares) compared to a reference run without atom (red points). Within the errors, no deviation in the pulse shape is observable, except for a small amplitude change stemming from the slightly different reflectivities, see Fig.\,\ref{fig:IndividualSteps}B.
(\textbf{C}) Nondestructive detection of a single photon. The probability of detecting zero or one photon is plotted. Yellow: Result of the SPCM detection. Grey: Calculated input pulse, taking into account the SPCM detection efficiency. Red: Result of the atomic state readout. Green: Atomic state readout without impinging light. Blue: Atomic state, conditioned on the SPCM detection of a reflected photon in the trigger interval.
}
\end{figure}

Having characterized the individual steps of the protocol, they are now combined to detect photons in a nondestructive way. The atom is prepared in the superposition state $\frac{1}{\sqrt{2}}(\ket{1}_a+\ket{2}_a)$. Within a $2.5$\,\textmu s long trigger interval, we send in a weak coherent laser pulse with an average photon number of $\bar{n}=0.115(11)$ and monitor its reflection with conventional single-photon counting modules (SPCMs). Fig.\,\ref{fig:QND}A shows a typical experimental run, where a photon was subsequently detected (red line in the blue trigger interval). Therefore, after $\pi/2$ rotation of the atomic state, many fluorescence photons are observed \cite{bochmann_lossless_2010} in the 25\,\textmu s long readout interval (grey), unambiguously signalling the atomic state change induced by the detected photon. Thus, in the case shown in Fig.\,\ref{fig:QND}A, a photon is detected twice: by the nondestructive detector and with a conventional, absorptive SPCM. The depicted trace also indicates that the setup works as an all-optical switch \cite{chen_all-optical_2013} which does not destroy the impinging trigger photon and also does not affect its temporal envelope. The latter can be seen in Fig.\,\ref{fig:QND}B, where the arrival-time histogram of the photons detected with the SPCMs after reflection from the setup is shown. The data taken during the nondestructive photon measurement (black squares) do not show a significant deviation from the reference curve recorded without atom (red points)---except for a small reduction in amplitude which is consistent with the results of Fig.\,\ref{fig:IndividualSteps}B.

When the input pulse is blocked, no photons are observed, neither in the blue nor in the grey interval of Fig.\,\ref{fig:QND}A, in $97.1(4)\%$ of all runs. In the remaining $2.9\%$, many fluorescence photons are observed during the atomic state readout, corresponding to a `dark count' of the nondestructive photon detector. This is caused by imperfections in the atomic state preparation, rotation and readout and might be improved by magnetic shielding of the setup and by using more complex state-rotation techniques such as composite pulses \cite{vandersypen_nmr_2005}.

We now investigate the photon detection efficiency of our nondestructive device. The probability of detecting a photon in the input pulse is given in Fig.\,\ref{fig:QND}C. The results obtained with calibrated conventional SPCMs, without and with correction for their limited quantum efficiency of $55(5)\%$, are shown as yellow and grey bars, respectively. The red bars are obtained from the atomic-state readout. Comparison of the grey and red bars shows good agreement, but does not reveal information about potential systematic errors. Therefore, we also analyze correlations between the detection of a reflected photon by the SPCMs and by our nondestructive detector. The blue bars show the probability of finding the atom in $\ket{2}_a$, conditioned on the detection of a photon by the SPCMs. We obtain $82.1(1.7)\%$. Correcting for the influence of two-photon components in the input laser field (and SPCM dark counts) \cite{som}, the conditional detection efficiency of our device for single photons is 87\%.

There are two major experimental imperfections \cite{som} that contribute to the deviation of the conditional detection efficiency from unity. First, the spatial mode matching of the input photons and the cavity mode (92(2)\%; corresponding reduction 12(3)\%), and second, the fidelity of the atomic state preparation, rotation and readout (estimated reduction 3\%) \cite{som}. None of the imperfections has a fundamental limit. Therefore, it should be possible to further increase the efficiency achieved in our first proof-of-principle experiment, which already compares well with state-of-the-art absorbing single-photon detectors \cite{hadfield_single-photon_2009,eisaman_single_2011,marsili_detecting_2013}.

The probability that an impinging photon is reflected is on average 66(2)\%. If a photon is absorbed, the atomic state is projected and the detection process gives the wrong result with a probability of 50\%. Therefore, the probability to detect a single input photon without postselection on its reflection from the cavity is calculated to be 74\% \cite{som}.

In contrast to all absorbing detectors, the efficiency of our detector can be further improved by attempting more measurements. Concatenating two of our devices is expected to increase the detection efficiency to 87\%, while three or more devices should yield 89\% \cite{som}. The achieved value is currently limited by absorption and scattering losses of both the atom and the cavity mirrors. To further improve, a decrease in cavity loss or an increase in atom-cavity coupling strength would be required. Both can be achieved either in Fabry-Perot \cite{colombe_strong_2007} or other \cite{Dayan_photon_2008,junge_strong_2013,thompson_coupling_2013} resonators.

The atom-photon interaction mechanism that has been presented in this work lays the ground for numerous experiments. A first step is the repeated nondestructive measurement of a single optical photon. Next, with a higher number of photons in the impinging laser pulse, the odd-even parity measurement allows one to generate new quantum states of optical light fields, e.g. Schr\"odinger-cat states \cite{wang_engineering_2005}. Measuring the phase of the reflected light could be used to entangle two atoms in the cavity \cite{soerensen_measurement_2003}. Moreover, using the polarization degree of freedom as a qubit should facilitate a deterministic quantum gate between a single photon and a single atom \cite{duan_scalable_2004, cho_generation_2005}. This can be further extended to an entangling gate between several successively impinging photons \cite{duan_scalable_2004} or between several atoms trapped in the same or even in remote cavities, thus efficiently generating atomic cluster states \cite{xiao_realizing_2004,cho_generation_2005,duan_robust_2005}. Implementing this gate operation would also allow for a deterministic photonic Bell-state measurement, which would increase the efficiency of measurement-based quantum networks with remote single atoms \cite{moehring_entanglement_2007,nolleke_efficient_2013} close to unity.

\begin{acknowledgments}
We thank Norbert Kalb for experimental assistance. This work was supported by the European Union (Collaborative Project SIQS) and the Bundesministerium f\"ur Bildung und Forschung via IKT 2020 (QK\_QuOReP).
\end{acknowledgments}

\setcounter{equation}{0}
\renewcommand{\theequation}{S\arabic{equation}}
\section*{Supplementary Materials}
\section*{Materials and methods}
\subsection*{Atomic state preparation}
The level scheme of $^{87}$Rb exhibits eight ground states: three in the $F=1$ and five in the $F=2$ manifold of the $5^2S_{1/2}$ state. The states $\ket{1}_a$ and $\ket{2}_a$ mentioned in the main text are identified with the $\ket{F, m_F}$ states $\ket{1,1}$ and $\ket{2,2}$, respectively. The excited state $\ket{3}_a$ is the $\ket{3,3}$ state of the $5^2P_{3/2}$ manifold. The atomic transition $\ket{2}_a \leftrightarrow \ket{3}_a$ is Stark-shifted by 100\,MHz due to the trap light.

To prepare the atom in $\ket{2,2}$, a 250\,\textmu s long interval of optical pumping with a coherent laser pulse is used. This pump pulse is resonant with the cavity and the $\ket{2}_a \leftrightarrow \ket{3}_a$ transition, circularly polarized and applied through the highly reflecting cavity mirror. A repumping laser on the $F=1 \leftrightarrow F'=2$ transition of the D$_2$ line is applied perpendicular to the cavity axis. When the atom is in $\ket{2,2}$, the transmission of the pump laser through the cavity is reduced by a factor of ten \cite{reiserer_ground-state_2013}. This allows to detect whether the atom has been pumped to the desired state by measuring the transmitted pump light with SPCMs. Figure \ref{fig:StateDetection}A shows the distribution of the number of photons detected in a 10\,\textmu s interval. At the beginning of the optical pumping process (black squares), the atom is typically in a state that does not couple to the cavity. A Poissonian distribution (black fit curve) with an average of 17 detected photons is observed. At the end of the pumping process (red points), the photon number distribution is again Poissonian (red fit curve, 1.6 detected photons on average), however clearly distinct from the high photon numbers observed for the uncoupled atom. During optical pumping, one can directly observe the increase of the atomic population in the desired state as a developing peak around low photon numbers (blue triangles). At the end of the pumping interval, still more than five transmitted photons are observed in 13\% of all cases. We associate these events with the atom not having been pumped to $\ket{2,2}$ and therefore exclude them in the analysis of the data. In other words, we use our nondestructive single-photon detector only when the state preparation has been successful with high probability.

\renewcommand{\thefigure}{S1}
\begin{figure}[h!]
\includegraphics[width=1.0\columnwidth]{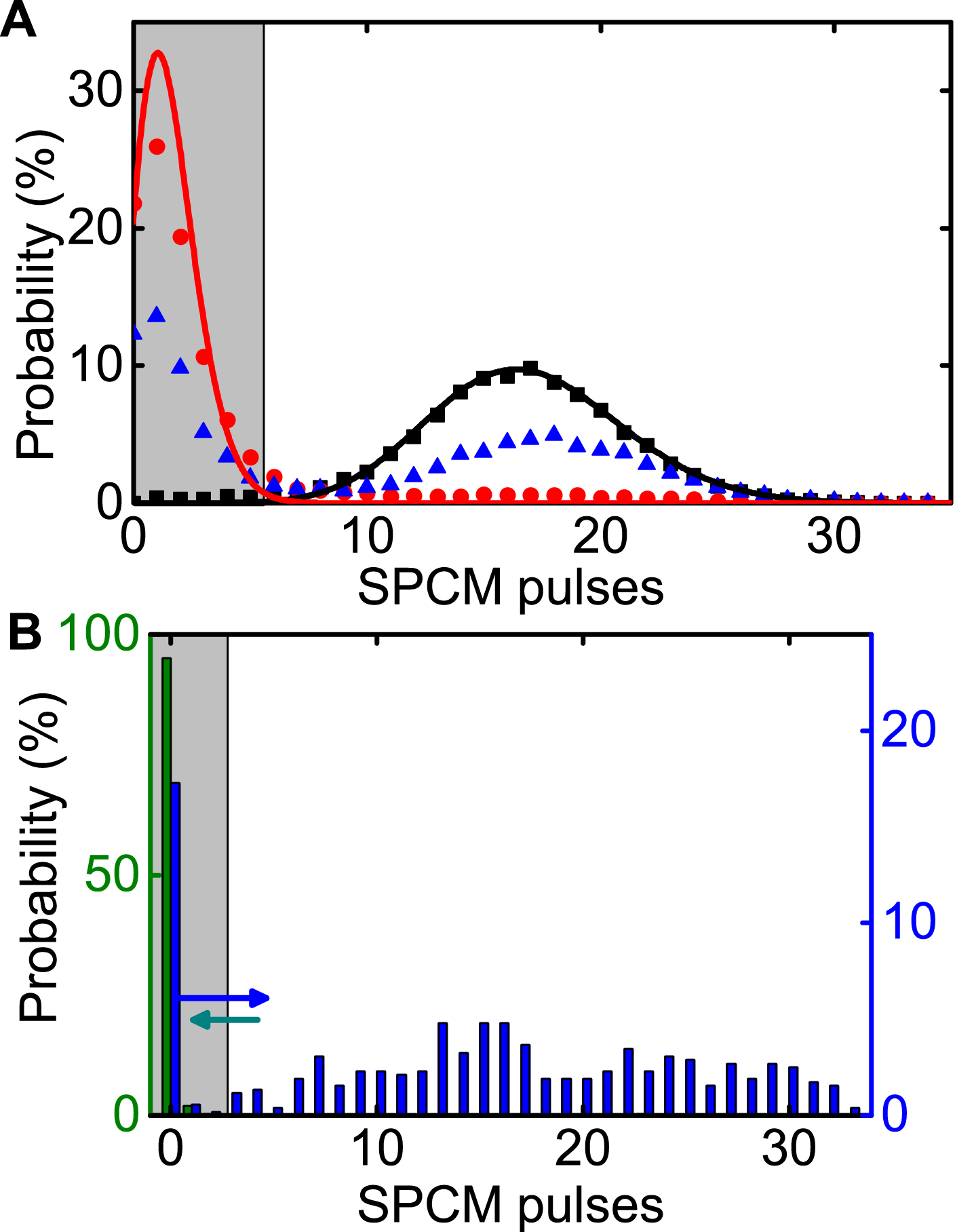}
\caption{\label{fig:StateDetection}
\textbf{Atomic state preparation and detection.} (\textbf{A}) Cavity transmission during optical pumping. At the beginning of the pumping process (first 10\,\textmu s, black squares), we detect a Poissonian distribution (black fit curve) of photon numbers with 16 detected photons on average. At the end of the pumping process, we only detect 1.6 photons on average (red points and red Poissonian fit curve). In between, a steady increase of the atomic population in $\ket{2,2}$ can be seen (blue triangles). When 5 or less pulses are detected (grey box, 87\% of all experimental runs), we infer that the atom was correctly pumped to the state $\ket{2,2}$. (\textbf{B}) State-detection measurement when the input pulse is blocked (green, left axis) or when a photon has been detected by the SPCM during the trigger interval (blue, right axis). The discrimination threshold for state detection is chosen to be 2.5 SPCM pulses, as indicated by the grey area.
}
\end{figure}

\subsection*{Atomic state rotation}
In order to implement the atomic state rotation, a pair of copropagating Raman beams applied perpendicular to the cavity axis is used \cite{reiserer_ground-state_2013}, 0.1\,THz red-detuned from the D$_1$ line at 795\,nm. A magnetic field is applied to split the ground-state Zeeman states with $\Delta m_F =1$ by 0.4\,MHz, such that the transition $\ket{2,2} \leftrightarrow \ket{1,1}$ can be addressed in a frequency-selective way without driving transitions to other Zeeman states. Thus, the Raman beams lead to an effective two-level coupling between the states $\ket{2}_a$ and $\ket{1}_a$.

\subsection*{Atomic state readout}
For single-shot readout of the atomic state, a 25\,\textmu s long laser pulse is applied from the side of the cavity in a counterpropagating configuration with orthogonal linear polarizations. The laser is resonant with the cavity and the Stark-shifted atomic transition $\ket{2}_a \leftrightarrow \ket{3}_a$. If the atom is in $F=1$ and therefore off-resonant, it does not scatter any photons. If it is in $F=2$, however, we observe Purcell-enhanced scattering into the cavity mode. This allows to discriminate the atomic state with high fidelity \cite{bochmann_lossless_2010}. Figure \ref{fig:StateDetection}B shows a histogram of the number of photons detected with the SPCM for the same data set as in Fig.\,\ref{fig:QND}B. When the input pulse is blocked, the atom ideally ends up in the off-resonant state $\ket{1}_a$ and less than three photons, caused by stray light and SPCM dark counts, are detected with a probability of $97.1(4)\%$ (green bars). When a photon has been detected by the SPCM during the trigger interval, the atom is most likely transferred to $\ket{2}_a$, and three or more photons are detected with a probability of $82.1(1.7)\%$ (blue bars).

\subsection*{Calculation of the single-photon detection efficiency}
To characterize our device, we measure the probability that the atom is found in $\ket{2}_a$, conditioned on an SPCM detection event to be $\eta_\mathrm{cond}=82.1(1.7)\%$. In this measurement, weak coherent laser pulses with an average photon number of $\bar{n}=0.115(11)$ are used, such that $p_1=10.3\%$ of the pulses contain a single photon, $p_2=0.6\%$ contain two and only a negligible fraction of $0.02\%$ contains more than two. Input states with two photons are ideally not detected by our setup, because the resulting phase shift of $2\pi$ returns the atom to its original state. Therefore, characterization of the device with coherent pulses yields a conditional detection efficiency which is systematically lower than $\eta_\mathrm{cond}^\mathrm{n=1}$, the value expected for single-photon input pulses. Because of the low probability for more than two photons in the incoming pulse, they have a negligible influence and are therefore neglected in the following. The detection of exactly one photon by the SPCM can be the result of four different scenarios:

\begin{itemize}
\item
The most likely, ideal situation is that the impinging pulse contains one photon, which is detected by the SPCM with a probability of $p_\mathrm{det}= r \epsilon$.  Here, $r=0.66(2)$ is the probability that the photon is reflected from the cavity and $\epsilon=0.55(5)$ is the quantum efficiency of the SPCM. The total probability for this case is $p_1 p_\mathrm{det}$.
\item
In the second case, there are two photons in the impinging pulse, both of which are reflected off the cavity, but only one is detected by the SPCM. This happens with a probability of $p_2 p_\mathrm{det,refl}$, with $p_\mathrm{det,refl} = 2 r^2 \epsilon (1 - \epsilon)$. In this case, the atomic phase shift is $2\pi$ and our nondestructive detection mechanism yields the incorrect result ``no photon''.
\item
In the third case, one of the two photons is detected, but the other is absorbed or scattered by our setup. This happens with a probability of $p_2 p_\mathrm{det,abs}$, with $p_\mathrm{det,abs}= 2 r \epsilon (1-r)$, and projects the atom to either state $\ket{1}_a$ or state $\ket{2}_a$. The final $\pi/2$ rotation then results in an equal superposition of $\ket{1}_a$ and $\ket{2}_a$, such that our nondestructive detection mechanism leads to the incorrect result ``no photon'' with $50\%$ probability.
\item
Finally, a small fraction of 0.4\% of the SPCM detection events are caused by stray or dark counts without any impinging trigger photon. The corresponding probability for a dark count is $p_\mathrm{dark} = 1.6 \times 10^{-4}$. Since there is no phase shift on the atom, our detector will give the result ``no photon''.
\end{itemize}
Considering these cases, the conditional detection efficiency of our device for single-photon input pulses is:
$$\eta_\mathrm{cond}^\mathrm{n=1} = \frac{p_\mathrm{tot} \eta_\mathrm{cond} - \frac{1}{2} p_2 p_\mathrm{det,abs}}{p_1 p_\mathrm{det}} = 87\%.$$
Here, $p_\mathrm{tot} = p_\mathrm{dark} + p_1 p_\mathrm{det} + p_2 (p_\mathrm{det,refl} + p_\mathrm{det,abs})$ is the total probability for a single SPCM detection event.

To derive the \emph{unconditional} single-photon detection efficiency $\eta$, two cases have to be considered: In the first case, a single impinging photon is reflected and the detection mechanism works as intended. In the second case, a single photon is impinging, but not reflected and the atomic state is thereby projected: If the photon is transmitted through the high-reflector or absorbed or scattered by the mirrors, it first had to enter the cavity, which means that the atom is in $\ket{1}_a$ with a high probability. If the photon is scattered by the atom, the latter ends up in $\ket{2_a}$. In both cases, atomic state rotation and subsequent readout give the correct result only with a probability of $50\%$. Therefore, we calculate the unconditional single-photon detection efficiency of our device to be
$$\eta = r \eta_\mathrm{cond}^\mathrm{n=1} + (1-r)\frac{1}{2} = 74\%.$$

\subsection*{Detection efficiency when concatenating several devices}
For $m$ concatenated devices, the efficiency to detect a single photon increases to
$$\eta \sum_{i=0}^{m-1} \left(r \left(1 - \eta \right)\right)^i.$$
This yields 87\% for two of our nondestructive detectors, and 89\% for three or more devices.

\subsection*{Analysis of experimental imperfections}
In addition to the influence of coherent input pulses and dark counts discussed earlier, several experimental imperfections contribute to the reduction of the detection efficiency:
\begin{itemize}
\item
The spatial mode-matching of the input photons and the cavity mode is not perfect (estimated to be $q=92(2)\%$). Therefore, some photons do not interact with the cavity. They are perfectly reflected and reach the SPCM with unit efficiency, but do not leave a trace in the atom. The photons interacting with the cavity, on the other hand, only reach the detector with probability $r$. Therefore, $\frac{1-q}{1-q+r q}=12(3)\%$ of the photons seen by the SPCM did not interact with the cavity.
\item
The atomic state preparation, rotation and readout are not perfect. We estimate a corresponding reduction in detection efficiency by 3\%.
\item
The reflection probabilities for the atom in state $\ket{1}_a$ (70\%) and $\ket{2}_a$ (62\%) are different. Therefore, reflection of the photon leaves the atom in a superposition state which is not exactly on the equator of the Bloch sphere. The estimated reduction is 0.4\%.
\item
Instability of the laser and cavity frequencies (standard deviation $\leq 300$\,kHz).
\item
Fluctuations of the atomic energy levels caused by light shifts.
\item
Cavity birefringence that can lead to noncircular polarization components which do not couple to the atomic state $\ket{3}_a$.
\end{itemize}
None of these imperfections has a fundamental limit, and we therefore expect that the efficiency of cavity-based nondestructive photon detection can be increased in future realizations.


\begin{thebibliography}{10}

\bibitem{glauber_quantum_1963}
R.~J. Glauber, The quantum theory of optical coherence. \emph{Phys. Rev.}
  \textbf{130}, 2529--2539 (1963).

\bibitem{mandel_optical_1995}
L.~Mandel, E.~Wolf, \emph{Optical Coherence and Quantum Optics} (Cambridge
  University Press, 1995).

\bibitem{braginsky_quantum_1996}
V.~B. Braginsky, F.~Y. Khalili, Quantum nondemolition measurements: the route from toys to tools. \emph{Rev. Mod. Phys.} \textbf{68}, 1--11 (1996).

\bibitem{grangier_quantum_1998}
P.~Grangier, J.~A. Levenson, J.-P. Poizat, Quantum non-demolition measurements
  in optics. \emph{Nature} \textbf{396}, 537--542 (1998).

\bibitem{wiseman_quantum_2010}
H.~M. Wiseman, G.~J. Milburn, \emph{Quantum Measurement and Control} (Cambridge
  University Press, 2009).

\bibitem{obrien_optical_2007}
J.~L. {O'Brien}, Optical quantum computing. \emph{Science} \textbf{318},
  1567--1570 (2007).

\bibitem{gisin_quantum_2007}
N.~Gisin, R.~Thew, Quantum communication. \emph{Nature Photon.} \textbf{1},
  165--171 (2007).

\bibitem{kimble_quantum_2008}
H.~J. Kimble, The quantum internet. \emph{Nature} \textbf{453}, 1023--1030
  (2008).

\bibitem{duan_scalable_2004}
L.-M. Duan, H.~J. Kimble, Scalable photonic quantum computation through
  cavity-assisted interactions. \emph{Phys. Rev. Lett.} \textbf{92}, 127902
  (2004).

\bibitem{nogues_seeing_1999}
G.~Nogues \emph{et~al.}, Seeing a single photon without destroying it.
  \emph{Nature} \textbf{400}, 239--242 (1999).

\bibitem{guerlin_progressive_2007}
C.~Guerlin \emph{et~al.}, Progressive field-state collapse and quantum
  non-demolition photon counting. \emph{Nature} \textbf{448}, 889--893 (2007).

\bibitem{johnson_quantum_2010}
B.~R. Johnson \emph{et~al.}, Quantum non-demolition detection of single
  microwave photons in a circuit. \emph{Nature Phys.} \textbf{6}, 663--667
  (2010).

\bibitem{cho_generation_2005}
J.~Cho, H.-W. Lee, Generation of atomic cluster states through the cavity
  input-output process. \emph{Phys. Rev. Lett.} \textbf{95}, 160501 (2005).

\bibitem{bochmann_lossless_2010}
J.~Bochmann \emph{et~al.}, Lossless state detection of single neutral atoms.
  \emph{Phys. Rev. Lett.} \textbf{104}, 203601 (2010).

\bibitem{som}
Materials and methods are available as supplementary material on \textit{Science}
  {O}nline.

\bibitem{reiserer_ground-state_2013}
A.~Reiserer, C.~N\"{o}lleke, S.~Ritter, G.~Rempe, Ground-state cooling of a
  single atom at the center of an optical cavity. \emph{Phys. Rev. Lett.}
  \textbf{110}, 223003 (2013).

\bibitem{walls_quantum_2008}
D.~F. Walls, G.~J. Milburn, \emph{Quantum Optics} (Springer, 2008).

\bibitem{chen_all-optical_2013}
W.~Chen \emph{et~al.}, All-optical switch and transistor gated by one stored
  photon. \emph{Science} \textbf{341}, 768--770 (2013).

\bibitem{vandersypen_nmr_2005}
L.~M.~K. Vandersypen, I.~L. Chuang, {NMR} techniques for quantum control and
  computation. \emph{Rev. Mod. Phys.} \textbf{76}, 1037--1069 (2005).

\bibitem{hadfield_single-photon_2009}
R.~H. Hadfield, Single-photon detectors for optical quantum information
  applications. \emph{Nature Photon.} \textbf{3}, 696--705 (2009).

\bibitem{eisaman_single_2011}
M.~D. Eisaman, J.~Fan, A.~Migdall, S.~V. Polyakov, Single-photon sources and
  detectors. \emph{Rev. Sci. Instrum.} \textbf{82}, 071101 (2011).

\bibitem{marsili_detecting_2013}
F.~Marsili \emph{et~al.}, Detecting single infrared photons with 93\% system
  efficiency. \emph{Nature Photon.} \textbf{7}, 210--214 (2013).

\bibitem{colombe_strong_2007}
Y.~Colombe \emph{et~al.}, Strong atom-field coupling for {Bose-Einstein}
  condensates in an optical cavity on a chip. \emph{Nature} \textbf{450},
  272--276 (2007).

\bibitem{Dayan_photon_2008}
B.~Dayan \emph{et~al.}, A photon turnstile dynamically regulated by one atom.
  \emph{Science} \textbf{319}, 1062--1065 (2008).

\bibitem{junge_strong_2013}
C.~Junge, D.~{O'Shea}, J.~Volz, A.~Rauschenbeutel, Strong coupling between
  single atoms and nontransversal photons. \emph{Phys. Rev. Lett.}
  \textbf{110}, 213604 (2013).

\bibitem{thompson_coupling_2013}
J.~D. Thompson \emph{et~al.}, Coupling a single trapped atom to a nanoscale
  optical cavity. \emph{Science} \textbf{340}, 1202--1205 (2013).

\bibitem{wang_engineering_2005}
B.~Wang, L.-M. Duan, Engineering superpositions of coherent states in coherent
  optical pulses through cavity-assisted interaction. \emph{Phys. Rev. A}
  \textbf{72}, 022320 (2005).

\bibitem{soerensen_measurement_2003}
A.~S. S{\o}rensen, K.~M{\o}lmer, Measurement induced entanglement and quantum
  computation with atoms in optical cavities. \emph{Phys. Rev. Lett.} \textbf{91}, 097905 (2003).

\bibitem{xiao_realizing_2004}
Y.-F. Xiao \emph{et~al.}, Realizing quantum controlled phase flip through
  cavity {QED}. \emph{Phys. Rev. A} \textbf{70}, 042314 (2004).

\bibitem{duan_robust_2005}
L.-M. Duan, B.~Wang, H.~J. Kimble, Robust quantum gates on neutral atoms with
  cavity-assisted photon scattering. \emph{Phys. Rev. A} \textbf{72}, 032333
  (2005).

\bibitem{moehring_entanglement_2007}
D.~L. Moehring \emph{et~al.}, Entanglement of single-atom quantum bits at a
  distance. \emph{Nature} \textbf{449}, 68--71 (2007).

\bibitem{nolleke_efficient_2013}
C.~N\"{o}lleke \emph{et~al.}, Efficient teleportation between remote
  single-atom quantum memories. \emph{Phys. Rev. Lett.} \textbf{110}, 140403
  (2013).

\end{thebibliography}
\end{document}